\documentclass[10pt,journal]{IEEEtran}

\usepackage{cite}

\ifCLASSINFOpdf
\else
  \usepackage[dvips]{graphicx}
  \DeclareGraphicsExtensions{.eps}
\fi
\usepackage[cmex10]{amsmath}
\usepackage{amsfonts,amssymb,amsthm,mdwmath}
\newtheorem{remark}{Remark}
\newtheorem{lemma}{Lemma}

\usepackage{array}

\ifCLASSOPTIONcompsoc
  \usepackage[caption=false,font=normalsize,labelfont=sf,textfont=sf]{subfig}
\else
  \usepackage[caption=false,font=footnotesize]{subfig}
\fi

\usepackage{stfloats}
\usepackage{url}
\usepackage{acronym}

\usepackage[usenames, dvipsnames]{color}
\begin{document}
\acrodef{LoS}[LoS]{Line-of-Sight}
\acrodef{NLoS}[NLoS]{Non-Line-of-Sight}
\acrodef{RWPM}[RWPM]{Random Waypoint Mobility}
\acrodef{PDF}[PDF]{Probability Distribution Function}
\acrodef{RV}[RV]{Random Variable}
\acrodef{PPP}[PPP]{Poisson Point Process}
\acrodef{MGF}[MGF]{Moment Generating Function}
\acrodef{RWPM}[RWPM]{Random Waypoint Mobility}
\acrodef{i.i.d.}[i.i.d.]{independent and identically distributed}
\acrodef{1D}[1D]{one-dimensional}
\acrodef{2D}[2D]{two-dimensional}

\title{Temporal Correlation of Interference in Bounded Mobile Ad Hoc Networks with Blockage}
%
%
%


\author{Konstantinos Koufos and Carl P. Dettmann 
 \IEEEcompsocitemizethanks{\IEEEcompsocthanksitem The authors are with the School of Mathematics, University of Bristol, BS8 1TW, Bristol, UK. \{K.Koufos, Carl.Dettmann\}@bristol.ac.uk}} 
\maketitle

\begin{abstract}
In mobile wireless networks with blockage, different users, and/or a single user at different time slots, may be blocked by some common obstacles. Therefore the temporal correlation of interference does not depend only on the user displacement law but also on the spatial correlation introduced by the obstacles. In this letter, we show that in mobile networks with a high density of users, blockage increases the temporal correlation of interference, while in sparse networks blockage has the opposite effect. 

\end{abstract}


%
\IEEEpeerreviewmaketitle

\section{Introduction}

\IEEEPARstart{T}{HE} temporal correlation of interference affects the temporal correlation of outage, and subsequently, it impacts many network performance metrics, e.g., end-to-end throughput, multi-hop delay, etc. Assuming uncorrelated user activity and fading over time, the user mobility is the main factor reducing the temporal correlation of interference~\cite{Ganti2009, Gong2011, Schilcher2012, Gong2014}.

In areas with blockage, different users as well as a single user at different time slots, may be blocked by some common obstacles. The interference level is dominated by the \ac{LoS} transmissions, and the transitions between \ac{LoS} and \ac{NLoS} propagation conditions due to mobility will reduce the temporal correlation of interference. However, blockage simultaneously introduces   spatial correlation among the users. This is due to correlated penetration losses. Mobility, however, will not induce significant decorrelation of  interference when the number of users is high. Studying the impact of blockage on the moments of interference is a topic of growing interest~\cite{Bai2014,Bai2015,Thornburg2016}, considering the ongoing standardization activities for commercial wireless networks in  millimeter-wave bands. Nevertheless, interference correlation with blockage is yet to be studied. 


Without blockage, the temporal correlation of interference depends on the user displacement law~\cite{Koufos2016}. In this letter, we show that with blockage, the correlation of interference does not depend only on the mobility and the penetration losses but also on the user density. In sparse networks, where the spatial correlation among the users is negligible, the transitions in the propagation conditions from \ac{LoS} to \ac{NLoS} due to mobility dominate the temporal statistics of interference. As a result, blockage reduces the temporal correlation of interference. On the other hand, in dense networks, the correlation in the interference levels generated by different users dominates the temporal correlation of interference and mobility may not help much in reducing it. 

In our analysis, we use the \ac{RWPM} model, e.g.~\cite{Gorawski2014},  because it has some desirable features for our problem: It is defined over a finite area, and it results in a non-uniform distribution of users.  We use  the \ac{RWPM} model over a \ac{1D} lattice because in that case the user displacement law is known for time-lags equal to one and two time slots~\cite{Koufos2016}. For larger time-lags, approximations to the user displacement are also available for a zero think time. Note that by increasing the lattice size and at the same time the user speed, one can obtain  approximations for the continuous \ac{1D} space. Even though, a \ac{2D} deployment would be naturally more relevant, the \ac{1D} scenario allows to get analytical insight on the system behaviour. Also, one can still find practical applications, e.g., correlation of interference in vehicular networks. 
\section{System model}
\label{sec:System}
We consider a Poisson number of users, with mean $K$, which are moving across a \ac{1D} lattice of size $N$. Each user selects uniformly at random a destination, and travels with a constant speed $u$ lattice points per time slot. When it reaches the destination, it stops and thinks for a number of time slots selected from the discrete uniform distribution on $\left\{0,1,\ldots M\right\}$. Let us denote the \ac{RV} of the $i$-th user location by $x_i$. Its \ac{PDF} in the steady state is~\cite{Koufos2016}
\begin{equation}
\label{eq:SteadyPDF}
f_{x_i}(n) \!=\! \frac{p}{N} \!+\! \left(1\!-\!p\right) \!\frac{3N\left(2n\!-\!1\right)\!-\!6n\left(n\!-\!1\right)\!-\!3}{N(N^2-1)}, n\!\leq\! N
\end{equation}
where $p\!=\!\frac{M\!/2}{M\!/2+(N\!+\!1)\!/\left(3u\right)}$  is the average think time for a randomly selected user. 

Given the location $n$, let us denote by $\mathbb{P}({n\!+\!k},\tau)$ the probability that the user is located at the lattice point $(n\!+\!k)$ after $\tau$ time slots. The \ac{RWPM} model  introduces different levels of mobility at different locations. For instance, the  probability that a user thinks at the lattice point $n$ is $\mathbb{P}(n,1)\!=\!\frac{p}{Nf_{x}(n)}$, which means that the users close to the center tend to move with higher probability than the users near the boundaries~\cite{Koufos2016}. We compute the interference  at the locations, $y_p\!=\!n\! + c, n\!=\!1,2,\ldots,\left\lfloor\frac{N}{2}\right\rfloor$ and $c\in(0,1)$. 


Let consider a Poisson number of obstacles, with mean $N_{\!o}$, distributed uniformly at random in the continuous space $[1,N]$. The obstacles do not hinder the user moves, but they attenuate the user signal. The number of obstacles $n_{\!o}$ on the link $x_i\!\rightarrow\! y_p$, between the $i$-th user and the location $y_p$, is a Poisson \ac{RV} with parameter $q_iN_{\!o}$, $\text{Po}(q_iN_{\!o})$, where $q_i\!=\!\frac{d_i}{N-1}, d_i\!=\!|x_i\!-\!y_p|$.  The fraction of penetration power loss per obstacle follows the uniform distribution on $[0,\gamma]$, $\gamma\leq 1$. The fraction of penetration loss, $\beta_i$, over the link $x_i\! \rightarrow \!y_p$ is equal to the product of the power loss fractions from all obstacles on that link. Note that the \acp{RV} $\beta_i$ and $x_i$ are dependent, e.g., the longer the link $x_i\! \rightarrow \!y_p$ is, the higher the penetration loss should be, because more obstacles are likely to block the user.

Assuming common transmit power level $P_t$ for all users, the  interference at an arbitrarily selected time slot $t$ is 
\[
\mathcal{I}(t) = P_t \sum\nolimits_i {\xi_i(t)\, h_i(t) \, \beta_i(t)\, g\left(x_i(t)\!-\!y_p\right)}
\]
where $\xi_i$ is a Bernoulli \ac{RV} describing the $i$-th user activity, $\mathbb{E}\left\{\xi_i\right\}\!=\!\xi\, \forall i$, $h_i$ is an exponential \ac{RV} with unit mean modeling Rayleigh fading, $x_i\in \left\{1,2,\ldots, N\right\}$ is the \ac{RV} for the $i$-th user location with \ac{PDF} given in~\eqref{eq:SteadyPDF}, and  $g(x)=\frac{1}{\epsilon+|x|^a}$ is the distance-based propagation pathloss function, where $\epsilon$ is used to avoid singularity at $x\!=\!0$. 

It is assumed that the user activity and fading are \ac{i.i.d.} over time slots and users. On the other hand, with the \ac{RWPM} model, the locations of a user are correlated in time. Different users move independently of each other but their penetration losses are in general correlated because they may be blocked by some common obstacles. The \ac{MGF} of the interference at two time slots $t$ and $\tau$ is 
\[
\Phi_{\mathcal{I}} \!= \!\iint\limits\sum_{{ {\rm \xi, \rm x},i}}{e^{s_{\!1} \mathcal{I}(t) + s_2\mathcal{I}(\tau)}f_{\!\rm x,\boldsymbol \beta} \,f_{\xi}\, f_{{\rm h}} \, {\text{Po}}(K) {\rm d h d} \boldsymbol\beta}
\] 
where $\rm \xi$, $\rm h$, $\rm x$ and $\boldsymbol \beta$ are vectors of \acp{RV} with elements, $\xi_i, h_i$, $x_i$ and $\beta_i \, \forall i$ at time slots $t$ and $\tau$, ${\text{Po}}(K)\!=\!\frac{e^{-K}K^i}{i!}$, and the arguments in the \acp{PDF} are omitted for brevity. 

In order to describe the correlation of interference at time-lag $l\!=\!|t\!-\!\tau|$, we use the Pearson correlation coefficient which is defined as the ratio of the covariance of \acp{RV} $\mathcal{I}(t),\mathcal{I}(\tau)$ divided by the product of their standard deviations. In the steady state, the moments of interference become independent of the time we take the measurements, and the Pearson correlation coefficient  becomes 
\begin{equation}
\label{eq:CorrCoeff}
\rho_l = \frac{\mathbb{E}\left\{ \mathcal{I}(t)\, \mathcal{I}(\tau)\right\} - \mathbb{E}\left\{ \mathcal{I}(t) \right\}^2} {\mathbb{E}\left\{ \mathcal{I}^{\,2}(t) \right\} - \mathbb{E}\left\{ \mathcal{I}(t)\right\}^2}.
\end{equation}

\section{Interference mean and variance}
Conditioned on the number of obstacles $n_{\!o}\geq 1$ over the link $x_i\! \rightarrow \!y_p$, the \ac{PDF} of the fraction of penetration power loss  $f_{\beta_i|n_{\!o}}\triangleq h(\beta_{n_{\!o}})$ is equal to the \ac{PDF} of the product of $n_{\!o}$ \ac{i.i.d.} uniform \acp{RV} with support  $[0,\gamma]$. This \ac{PDF} is $h(\beta_{n_o}) \!=\! \frac{1}{\gamma^{n_o}(n_o-1)!} \left(\log\left(\frac{\gamma^{n_o}}{\beta_{n_o}}\right)\right)^{\!n_o\!-\!1}$ over the interval $[0,\gamma^{n_o}]$~\cite{Dettmann2009}.
The \ac{PDF} $f_{\beta_i}$ can be computed by averaging the \ac{PDF} $h(\beta_{n_{\!o}})$ over the Poisson \ac{RV} $n_{\!o}$. While it is difficult to express the \ac{PDF} $f_{\beta_i}$ in terms of simple functions, its moments can be computed as follows 
\begin{equation}
\label{eq:PDFbi}
\arraycolsep=1.4pt\def\arraystretch{2.2}
\begin{array}{lcl}
\mathbb{E}\left\{\beta_i^s\right\} &\stackrel{(a)}{=}& \displaystyle \int_{\beta_i}{\!\!\beta_i^s\sum\limits_{n_{\!o}=1}^\infty \!\!h(\beta_{n_{\!o}}) \text{Po}(q_iN_{\!o}) {\rm d}\beta_i} \!+\! e^{-q_iN_{\!o}}\\ 
 {} &\stackrel{(b)}{=}& e^{-q_iN_{\!o}\left(1-\frac{\gamma^s}{1+s}\right)} = e^{-\alpha d_i \left(1-\frac{\gamma^s}{1+s}\right)}.
\end{array}
\end{equation}

In (a), the rightmost term $e^{-q_iN_{\!o}}$ corresponds to the \ac{LoS} probability, i.e., $n_{\!o}\!=\!0$. In (b), we changed the orders of integration and summation because the \acp{RV} $\beta_{n_{\!o}}$ are independent of each other, we used that  $\mathbb{E}\left\{\beta_{n_{\!o}}^s\right\} \!=\!\gamma^{s n_o}(1\!+\!s)^{-n_o}$, and we averaged over the Poisson distribution $\text{Po}(q_iN_{\!o})$. The term $\alpha\!=\!\frac{N_{\!o}}{N\!-\!1}$ indicates the density of obstacles.

In order to compute the moments of interference, one has to average over the distributions of fading, penetration loss, number of users, user activity and location. 
\[
\arraycolsep=0.5pt\def\arraystretch{2.2}
\begin{array}{lcl}
\mathbb{E}\!\!\left\{\mathcal{I}\right\}\! &\stackrel{(a)}{=}& \displaystyle \sum\nolimits_i\!\mathbb{E}\!\left\{h_i\right\} \!\mathbb{E}\!\left\{\xi_i\right\} \!\!\sum\nolimits_{x_i}\!\! \int\limits{\!\!\!\beta_i g\!\left(d_{\!i}\right) \!f_{\beta_{\!i}|x_{\!i}} f_{x_{\!i}} {\rm d}\beta_{\!i}} \,{\text{Po}(\!K\!)} \\ 
&=& \displaystyle   \sum\nolimits_i \!\!\mathbb{E}\!\!\left\{h_i\right\} \mathbb{E}\!\!\left\{\xi_i\right\}\!\! \sum\nolimits_{n}{\mathbb{E}\!\left\{\beta_n\right\}g\!\left(d_{\!n}\right) f_{x_i}\!\!\left(n\right)} \,{\text{Po}(\!K\!)} \\
&\stackrel{(b)}{=}& K \xi \displaystyle \sum\nolimits_{n}{ e^{-\alpha d_n\left(1-\frac{\gamma}{2}\right)} g(d_n) f_x\!\left(n\right)}.
\end{array}
\]

In (a), we used that the \acp{RV} $x_i,\beta_i$, are dependent. In (b), we computed $\mathbb{E}\left\{\beta_n\right\}$ from equation~\eqref{eq:PDFbi} for $s\!=\!1$, we used that the users are indistinct, and we took the average in terms of the Poisson distribution $\text{Po}(\!K\!)$. The transmit power level has been taken equal to $P_t\!=\!1$ and $d_n\!=|\!n\!-\!y_p|$. Following the same assumptions, we get 
\[
\arraycolsep=0.5pt\def\arraystretch{2.5}
\begin{array}{lcl}
\mathbb{E}\!\left\{\mathcal{I}^{\,2}\right\} &=& \displaystyle 2 K \xi  \!\sum\nolimits_{n}{\!e^{-\alpha d_n\left(\!1-\frac{1}{3}\gamma^2\!\right)} \!g^2\!\!\left(d_{\!n}\right)\! f_x(n)} + K^2\xi^2\sigma
\end{array}
\]
where we used that $\mathbb{E}\!\left\{h_i^2\right\}\!=\!2$, $\mathbb{E}\!\left\{\xi_i^2\right\}\!=\!\xi$, $\mathbb{E}\left\{\beta_n^2\right\}\!=\!e^{-\alpha d_n\!\left(\!1\!-\!\frac{\gamma^2}{3}\!\right)}$, and   $\sigma\!=\!\sum\nolimits_{n,m}\!\mathbb{E}\!\left\{\beta_{\!n}\beta_{\!m}\right\} \! g\!\left(d_{n}\right)\! g\!\left(d_{m}\right)\! f_x(n) f_x(m)$ captures the correlation in the interference levels generated by different users. 
\begin{figure}[!t]
 \centering
  \includegraphics[width=3.5in]{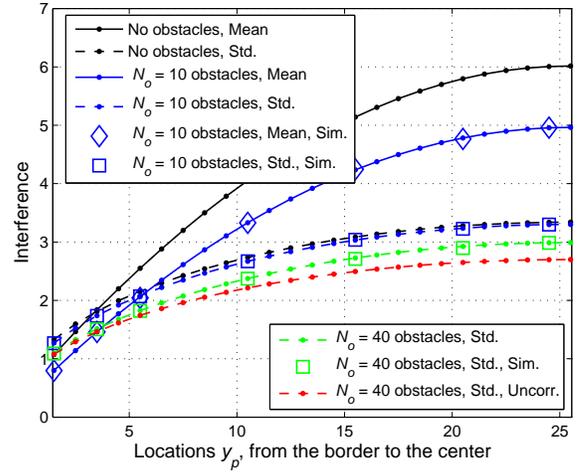}
 \caption{Mean and standard deviation of interference at the locations $y_{\!p}$ for $c\!=\!\frac{1}{2}$. The model is validated at seven locations for mean number of obstacles $N_{\!o}\!=\!10$ and $N_{\!o}\!=\!40$. The minimum attenuation per obstacle is $3$ dB or $\gamma\!=\!0.5$. Lattice size $N\!=\!50$, $K\!=\!50$ users, continuous user activity $\xi\!=\!1$, pathloss exponent $a\!=\!2$, $\epsilon\!=\!0.5$, and maximum think time  $M\!=\!5$ time slots.}
 \label{fig:MeanStd}
\end{figure}

In order to compute the cross-correlation of penetration loss we separate between the following cases: (i) $n\!>\!y_p$ and $m\!<\!y_p$ or $n\!<\!y_p$ and $m\!>\!y_p$. In that case, the links $n\!\rightarrow\! y_p$ and $m\!\rightarrow\! y_p$ do not share common obstacles and the penetration losses become uncorrelated. Thus,  $\mathbb{E}\!\left\{\beta_{\!n}\beta_{\!m}\right\}\!=\!  e^{-\alpha(d_n\!+\!d_m)\left(1\!-\!\frac{\gamma}{2}\right)}$. (ii) $n\!>\!y_p$ and $m\!>\!y_p$ or $n\!<\!y_p$ and $m\!<\!y_p$. Let assume that $d_m\!>\!d_n$. Then, $\mathbb{E}\!\left\{\beta_{\!n}\beta_{\!m}\right\}\!=\!\mathbb{E}\!\left\{\beta_{\!n}^2 \beta_{\!k}\right\}$, where $\beta_k$ is the penetration loss over the distance $d_{k}\!=\!d_{m}\!-\!d_{n}$. Since the penetration losses over the distances $d_n$ and $d_k$ are uncorrelated, $\mathbb{E}\!\left\{\beta_{\!n}\beta_{\!m}\right\}\!=\! e^{-\alpha\left( d_n\left(1\!-\!\frac{\gamma^2}{3}\right)\!+\!\left(d_m\!-\!d_n\right)\left(1\!-\!\frac{\gamma}{2}\right)\right)}$. Similarly, one can do the computation for $d_m\!\leq\!d_n$. Finally, $\mathbb{E}\!\left\{\beta_{\!n}\beta_{\!m}\right\}\!=\! e^{-\alpha\left( \min\left\{d_n,d_m\!\right\}\left(1\!-\!\frac{\gamma^2}{3}\right)\!+\!\left|d_m\!-\!d_n\right|\left(1\!-\!\frac{\gamma}{2}\right)\right)}$.
\begin{remark}
For impenetrable obstacles, $\gamma\!=\!0$, when $n\!>\!y_p$ and $m\!<\!y_p$ or vice-versa, $\mathbb{E}\!\left\{\beta_{\!n}\beta_{\!m}\right\}\!=\!e^{-\alpha(d_n\!+\!d_m)}$. Otherwise,  $\mathbb{E}\!\left\{\beta_{\!n}\beta_{\!m}\right\}\!=\!e^{-\alpha\max\left\{d_n,d_m\!\right\}}$.
\end{remark}

The calculation of the mean and standard deviation of interference are validated in Fig.~\ref{fig:MeanStd}. The impact of blockage on the mean is more prominent close to the center because over there, the term $e^{-\alpha d_n \left(1-\frac{\gamma}{2}\right)}$ filters out interference from both sides of location $y_{\!p}$. Near the boundaries, fewer users are located and the interference is practically generated from one direction. The standard deviation of the generated interference is affected less from blockage because: (i) The terms $e^{-\alpha d_n\left(1-\frac{1}{3}\gamma^2\right)}$ and $e^{-\alpha d_n \left(1-\frac{\gamma}{2}\right)}$, which are less than unity, are under the square root in the computation of the standard deviation. (ii) The correlation of   interference levels generated from different users increases the standard deviation. In  Fig.~\ref{fig:MeanStd}, one may see that by ignoring the spatial correlation, i.e., $\sigma\!=\! \frac{1}{K^2\sigma^2}\mathbb{E}\left\{\mathcal{I}\right\}^{\!2}$, the underestimation error may become non-negligilble. 

\section{Temporal interference correlation}
The cross-correlation of interference $\mathbb{E}\left\{I(t)I(\tau)\right\}$ depends on the user displacement law and the correlation of the \acp{RV} $\beta_i(t)$ and $\beta_j(\tau)$. After taking the first-order cross-derivative of the \ac{MGF} $\frac{\partial^2}{\partial s_1 \partial s_2} \Phi_{\mathcal{I}}\left(0,0\right)$, the cross-correlation at time-lag $l$ can be read as 
$\mathbb{E}\left\{ \mathcal{I}(t) \, \mathcal{I}(\tau)\right\} \!=\! \displaystyle K  \xi^2 \sigma_l  + K^2 \xi^2\sigma$, where  
\begin{equation}
\label{eq:Product}
\sigma_l\!=\!\! \sum\nolimits_{n,k}{\!\mathbb{E}\!\left\{ \beta_n \beta_{n+k}\right\}\!g(d_n)g(d_{n+k}) \mathbb{P}({n\!+\!k},\tau)f_x(n\!).}
\end{equation}
\begin{lemma}
Without blockage, i.e., $\alpha\!=\!0$, the correlation coefficient $\rho_l$ is independent of the user density. 
\begin{proof}
After replacing in equation~\eqref{eq:CorrCoeff}, $\sigma\!=\!\frac{1}{K^2\xi^2}\mathbb{E}\!\left\{\mathcal{I}\right\}^2$, we get 
$\rho_l|_{\alpha=0}=\frac{\xi \sum_{n,k} g(d_n)g(d_{n+k}) \mathbb{P}(n+k,\tau)f_x(n)}{2\sum_ng^2(d_n)f_x(n).}$, which is independent of the number of users $K$. Also, $\rho_l|_{\alpha=0}\leq\frac{\xi}{2}$. 
\end{proof}
\end{lemma}
\begin{lemma}
With blockage, the correlation coefficient $\rho_l$ increases with the number of users $K$.
\begin{proof} 
With blockage,  $\rho_l= \frac{c_1\!+\!c_2 K}{c_3\!+\!c_2 K }$ for $K\!\geq\! 2$ where $c_1\!=\!\xi\sigma_l$, $c_2\!=\!\xi\sigma\!-\!\xi \left( \sum\nolimits_n\mathbb{E}\left\{\beta_n\right\}g(d_n)f_x(n)\right)^2$, and $c_3\!=\!2\sum\nolimits_n\mathbb{E}\!\left\{\beta_n^2\right\}g^2(d_n)f_x(n)$, For  $K\!=\!1$, $\rho_l\!=\!\frac{c_1}{c_3}$. Since the terms $c_1, c_2, c_3$ are positive, and the Pearson correlation coeffcient is at most equal to unity, we get  $c_3\geq c_1$. Based on that, we can show that the derivative of $\rho_l$ in terms of $K$ is positive.
\end{proof}
\end{lemma}
\begin{remark}
If we expand $\rho_l$ around $K\!\rightarrow\!\infty$, we get $\rho_l\!=\!1\!-\!\frac{c_3-c_1}{K c_2}\!+\!\mathcal{O}\left(\frac{1}{K}\right)^2$. Therefore, by making the number of users $K$ sufficiently large, we can guarante that $\rho_l\!>\!\frac{\xi}{2}\geq\rho_l|_{\alpha=0}$. Using that $\mathbb{E}\left\{\beta_n\beta_{n+k}\right\}\!\leq\!\mathbb{E}\left\{\beta_n^2\right\} \forall \left\{n,k\right\}$, one can show that for $K\!=\!1$, $\rho_l\!=\!\frac{c_1}{c_3} \!\leq\! \rho_l|_{\alpha=0}$. Since $\rho_l$ increases with $K$ according to Lemma 2, there will be a critical number of users $K^*$ such that for $K>K^*$ we have $\rho_l>\rho_l|_{\alpha=0}$. The critical number $K^*$ is different at different points of the lattice.
\end{remark}

Equation~\eqref{eq:Product} can be used to calculate the correlation of interference for any mobility model. Next, we show how to compute the correlation, $\mathbb{E}\left\{ \beta_n \beta_{n+k}\right\}$, for time-lag $l\!=\!1$ and user speed $u\!=\!1$ under \ac{RWPM}. The number of obstacles over the link $n\!\!\rightarrow \! y_p$ follows the distribution $\text{Po}(\alpha d_n)$, and it remains to identify the distribution of obstacles over the link $(n\!+\!k)\!\!\rightarrow \! y_p$ for all possible displacements $k\in\left\{-1,0,1\right\}$. For that, we separate between four cases.  

\textbf{Case 1}: $n\!<\!\left\lfloor{y_{\!p}}\right\rfloor$. (i) If the user thinks with probability $\mathbb{P}(n,1)$, the \acp{RV} $\beta_n$ and $\beta_{n+k}$ are fully correlated. Hence, $\mathbb{E}\left\{\beta_n^2\right\} = e^{-\alpha d_n\left(1 - \frac{1}{3}\gamma^2\right)}$. (ii) If the user moves to the right with probability $\mathbb{P}({n\!+\!1},1)$, the number of obstacles that the user bypasses follows the Poisson distribution  ${\text{Po}}(\alpha)$. Hence, $\mathbb{E}\!\left\{\!\beta_n\beta_{n\!+\!1}\!\right\}\! = e^{-\alpha \left(d_n\!-\!1\right)\left( 1\!-\!\frac{1}{3}\gamma^2\right)} \, e^{-\alpha \left(1\!-\!\frac{\gamma}{2}\right)}$. (iii) If the user moves to the left with probability $\mathbb{P}({n\!-\!1},1)$, the extra number of obstacles blocking the user signal follows the Poisson distribution  ${\text{Po}}(\alpha)$ and, $\mathbb{E}\!\left\{\!\beta_n\beta_{n-1}\!\right\}\! = e^{-\alpha d_n\left( 1-\frac{1}{3}\gamma^2\right)} \, e^{- \alpha \left(1-\frac{\gamma}{2}\right)}$. Therefore for $n\!<n_{\!1}, n_{\!1}\!=\!\left\lfloor{y_p}\right\rfloor$, the term  $\sigma_{11}\triangleq\sigma_1|_{n<n_{\!1}}$, is 
\[
\arraycolsep=0.5pt\def\arraystretch{2.2}
\begin{array}{lcl}
\sigma_{11}&=&\displaystyle \sum\nolimits_{n\!=\!1}^{n_{\!1}\!-\!1} \! g(d_n) f_x(n) e^{-(1-\frac{1}{3}\gamma^2)\alpha d_n} \Big( \mathbb{P}(n,1)g(d_n)+  \\ & &  \!\!\!\!\!\!\!\!\!\! e^{\alpha(\frac{\gamma}{2}-\frac{\gamma^2}{3})} \mathbb{P}({n\!+\!1},\!1)g(d_{n\!+\!1})\!+\!e^{-\alpha (1\!-\!\frac{\gamma}{2})}  \mathbb{P}({n\!-\!1},\!1)g(d_{n\!-\!1}) \Big).
\end{array}\]

\textbf{Case 2:} $n\!>\!n_2,n_2\!=\!\left\lceil{y_p}\right\rceil$. Similar to Case 1, we may compute $\sigma_{12}\triangleq\sigma_1|_{n>n_{2}}$ 
\[
\arraycolsep=0.5pt\def\arraystretch{2.2}
\begin{array}{lcl}
\sigma_{12}&=&\displaystyle \sum\nolimits_{n=n_2\!+\!1}^{N} \!\! g(d_n) f_x(n) e^{-(1-\frac{1}{3}\gamma^2)\alpha d_n} \Big( \mathbb{P}(n,1)g(d_n)+  \\ & &  \!\!\!\!\!\!\!\!\!\! e^{\alpha(\frac{\gamma}{2}-\frac{\gamma^2}{3})} \mathbb{P}({n\!-\!1},\!1)g(d_{n\!-\!1})\!+\!e^{-\alpha (1\!-\!\frac{\gamma}{2})}  \mathbb{P}({n\!+\!1},\!1)g(d_{n\!+\!1}) \Big).
\end{array}\]

\textbf{Case 3:} $n\!=\!n_{\!1}$. When the user is located at $n_{\!1}\!\!=\!\left\lfloor{y_p}\right\rfloor$, $d_{n_{\!1}}\!=\!c$, and it moves to the left, $\mathbb{E}\!\left\{\!\beta_{n_{\!1}}\beta_{n_{\!1}-1}\!\right\}\! \!=\! e^{-\alpha c \left( 1\!-\!\frac{1}{3}\gamma^2\right)} \, e^{- \alpha \left(1-\frac{\gamma}{2}\right)}$. When it moves to the right, it passes over the location $y_{\!p}$ and the number of obstacles it sees at the two time slots are \ac{i.i.d.} Poisson \acp{RV}. Therefore $\mathbb{E}\left\{\beta_{n_{\!1}} \beta_{n_{\!1}+1}\right\}\!=\!e^{-\alpha c \left(1-\frac{\gamma}{2}\right)} e^{-\alpha \bar{c} \left(1-\frac{\gamma}{2}\right)} \!=\!e^{-\alpha  \left(1-\frac{\gamma}{2}\right)}$ where $\bar{c}\!=\!1\!-\!c$. The term $\sigma_{13}\triangleq\sigma_1|_{n=n_{\!1}}$, can be written as  
\[
\arraycolsep=1.4pt\def\arraystretch{2.2}
\begin{array}{lcl}
\sigma_{13} &=& g\left(c\right)f_x({n_{\!1}})e^{-\alpha c \left(1-\frac{1}{3}\gamma^2\right)} \Big( \mathbb{P}({n_{\!1}},\!1)g\!\left(c\right) \!+\! \\ 
& & e^{-\alpha \left(1-\frac{\gamma}{2}\right)} e^{\alpha c \left(1-\frac{1}{3}\gamma^2\right)} \mathbb{P}({n_{\!1}+1},\!1)g\!\left(\bar{c} \right)  \!+\! \\ 
& & e^{-\alpha \left(1-\frac{\gamma}{2}\right)} \mathbb{P}({n_{\!1}-1},\!1)g\!\left(1\!+\!c\right) \Big).
\end{array}
\]

\textbf{Case 4:} $n\!=\!n_{2}$. Similar to Case 3, the term  $\sigma_{14}\triangleq\sigma_1|_{n=n_{\!2}}, n_{2}\!=\!\left\lceil{y_p}\right\rceil$, and $d_{n_{2}}\!=\!\bar{c}$ can be written as  
\[
\arraycolsep=1.4pt\def\arraystretch{2.2}
\begin{array}{lcl}
\sigma_{14} &=& g\left(\bar{c}\right)\!f_x({n_2})e^{\!-\alpha \bar{c}  \left(1\!-\!\frac{1}{3}\gamma^2\right)} \Big( \mathbb{P}({n_{2}},\!1)g\!\left(\bar{c} \right) \!+\! \\ 
& &   e^{-\alpha \left(1-\frac{\gamma}{2}\right)} \mathbb{P}({n_2\!+\!1},\!1)g\!\left(1\!+\!\bar{c}\right)  \!+\! \\ & & e^{-\alpha \left(1-\frac{\gamma}{2}\right)} 
e^{\alpha \bar{c} \left(1-\frac{1}{3}\gamma^2\right)}
\mathbb{P}({n_2\!-\!1},\!1)g\!\left(c\right)   \Big).
\end{array}
\]
For impenetrable obstacles, which is a reasonable approximation for propagation in the millimeter-wave bands, the above equations are further simplified. Finally, one has to sum up the terms $\sigma_{1j},j\!=\!1,\ldots 4$, and the calculation of $\sigma_l$ for $l\!=\!1$ and $u\!=\!1$ is complete. The calculations for $l\!>\!1$ and $u\!>\!1$ can be carried out in a similar manner. 
\begin{figure}[!t]
 \centering
  \includegraphics[width=3.5in]{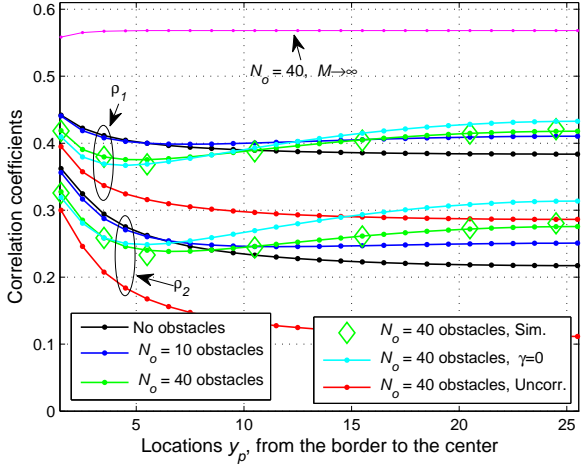}
 \caption{Correlation coefficients $\rho_1,\rho_2$ at the locations $y_{\!p}$.   The parameter settings are available in the caption of Fig.~\ref{fig:MeanStd} unless otherwise stated in the legend.}
 \label{fig:TempCorr}
\end{figure}


The correlation coefficients $\rho_1, \rho_2$ are depicted in Fig.~\ref{fig:TempCorr}. Without blockage, the temporal correlation of interference is higher close to the border because over there the level of mobility is lower. The impact of blockage on the correlation depends on the location and the density of users. Close to the boundaries, where the user density is low, the transitions from \ac{LoS} to \ac{NLoS} and vice versa dominate, and the interference correlation becomes less as compared to the case without obstacles. On the other hand, close to the center, where the user density increases, the correlated interference levels from the different users dominate over the randomness introduced by the mobility, and the correlation coefficients become higher. For the parameter settings used to generated Fig.~\ref{fig:TempCorr}, we observe cross-over points at some locations, see  Remark 2. One may also see that ignoring the correlated interference levels among the users results in significant underestimation errors for the correlation coefficients. Finally, in the limit of infinite think time, $M\!\rightarrow\!\infty$, the network becomes static and the user distribution uniform. Without blockage, the correlation coefficient  under Rayleigh fading and continuous user activity is equal to $\frac{1}{2}$~\cite{Gong2014}. In Fig.~\ref{fig:TempCorr}, we see that  blockage increases further the  correlation coefficient and also makes it location-dependent.  

\begin{figure}[!t]
 \centering
  \includegraphics[width=3.5in]{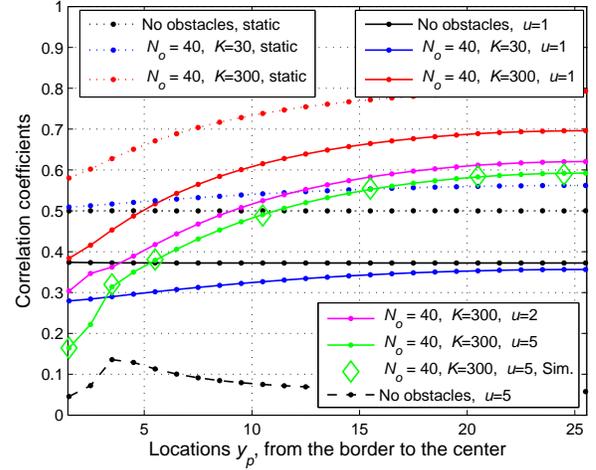}
 \caption{Correlation coefficients $\rho_1$ at the locations $y_{\!p}$ for different number of users $K$ and user speeds $u$. Think time $M=0$. The rest of the parameter settings are available in the caption of Fig.~\ref{fig:MeanStd}.}
\label{fig:TempCorr2}
\end{figure}

In Fig.~\ref{fig:TempCorr2}, we compare the correlation coefficients $\rho_1$ between a mobile network and a static network with user distribution given in~\eqref{eq:SteadyPDF}. Blockage increases the correlation coefficient $\rho_1$ in the static case, but mobility brings the correlation down when the number of users is low, e.g., $K\!=\!30$. When the user density is high, e.g., $K\!=\!300$, the spatial correlation among the users dominates, and mobility cannot make the correlation less than in the case without  blockage. For higher user speeds, $u\!=\!2, u\!=\!5$, the correlation remains high in the center where the user density is high. Close to the boundary, the high mobility along with the lower user density can make the correlation of  interference low but not less as compared to the case without blockage. 

\section{Conclusions}
In this letter, it is shown that correlated slow fading due to blockage can have a major impact on the temporal interference statistics. In the future, it is important to study in more detail the inter-play between user distribution, blockage distribution, mobility pattern and interference correlation also in \ac{2D} deployments. 

\begin{thebibliography}{1}
\bibitem{Ganti2009} R.~Ganti and M.~Haenggi, ``Spatial and Temporal Correlation of the Interference in ALOHA Ad Hoc Networks'', \emph{IEEE Commun. Lett.}, vol.~13, pp.~631-633, Sept.~2009.

\bibitem{Gong2011} Z.~Gong and M.~Haenggi, ``Temporal Correlation of the Interference in Mobile Random Networks'', \emph{IEEE Int. Conf. Commun. (ICC)},
Jun.~2011, pp.~1-5.

\bibitem{Schilcher2012} U.~Schilcher, C.~Bettstetter and G.~Brandner, ``Temporal correlation of interference in wireless networks with Rayleigh block fading'', \emph{IEEE Trans. Mobile Comput.}, vol.~11, pp.~2109-2120, Dec.~2012.

\bibitem{Gong2014} Z.~Gong and M.~Haenggi, ``Interference and Outage in Mobile
Random Networks: Expectation Distribution and Correlation'', \emph{IEEE
Trans. Mobile Comput.}, vol.~13, pp.~337-349, Feb.~2014.

\bibitem{Bai2014} T.~Bai, R.~Vaze, and R.W.~Heath, ``Analysis of Blockage Effects on Urban Cellular Networks'', \emph{IEEE Trans. Wireless Commun.}, vol.~13, pp.~5070-5083, 2014. 

\bibitem{Bai2015}  T.~Bai and R.W.~Heath,  ``Coverage and Rate Analysis for Millimiter-Wave Cellular Networks'', \emph{IEEE Trans. Wireless Commun.}, vol.~14, pp.~1100-1114, 2015.

\bibitem{Thornburg2016} A.~Thornburg, T.~Bai and R.W.~Heath, ``Performance Analysis of mmWave Ad Hoc Networks'', \emph{submitted for publication}, available at~\url{http://arxiv.org/abs/1412.0765}.

\bibitem{Koufos2016} K.~Koufos and C.P.~Dettmann, ``Temporal Correlation of Interference and Outage in Mobile Networks With Correlated Mobility in Finite Regions'', \emph{submitted for publication}, available at~\url{http://www.maths.bris.ac.uk/~macpd/sen/TempCorrInterf.pdf}.

\bibitem{Gorawski2014} M.~Gorawski, and K.~Grochla, ``Review of Mobility Models for Performance Evaluation of Wireless Networks'', Man-Machine Interactions 3, vol.~242, pp.~567-577, 2014. 

\bibitem{Dettmann2009} C.P.~Dettmann and O.~Georgiou, ``Product of $n$ Independent Uniform Random Variables'', \emph{Stat. and Probability Lett.,} vol.~79, no.~24, pp.~2501-2503, 2009.

\end{thebibliography}
\end{document}